# A Study of Experiment-based Radio Frequency Electromagnetic Field Exposure Evidence on Stochastic Nature of A Massive MIMO System

Tian Hong Loh[1], David Cheadle[1], Fabien Heliot[2], Ayodeji Sunday[3], Micheal Dieudonne[3]
[1] Electromagnetic & Electrochemical Technologies Department, National Physical Laboratory, Teddington, United Kingdom, tian.loh@npl.co.uk, david.cheadle@npl.co.uk
[2] Institute for Communication Systems, University of Surrey, Guildford, United Kingdom, f.heliot@surrey.ac.uk
[3] Keysight Technologies, ayodeji.sunday@keysight.com, michael_dieudonne@keysight.com

*Abstract*—In this paper, a massive multiple-input-multiple-output (mMIMO) testbed that is capable of mimicking realistic 5G new radio (NR) base station (BS) beamforming performance has been utilised to gather experimental-based evidence of 5G BS RF-EMF exposure within a real-world indoor environment. The mMIMO testbed has up to 128 RF channels with user-programmable software defined radio (SDR) capability. The stochastic nature of the 5G NR mMIMO system has been statistically assessed by evaluating the spatial variation of the RF-EMF exposure surrounding the mMIMO testbed when taking into account different beam profiles and data rates. Several other factors that influence the RF-EMF of mMIMO system have also being considered.

*Index Terms*—exposure, radiofrequency, electromagnetic field, massive mimo, stochastic nature, measurements.

## I. Introduction

The demand for high-speed communication for a range of diverse applications has driven the global digital agenda to better exploit the fifth generation (5G) new radio (NR) wireless technologies for fostering innovation and economic growth. 5G is currently being rolled out in a number of countries. However, public concern over potential health risks from radiofrequency electromagnetic field (RF-EMF) exposure from base stations (BSs) has led to stringent RF exposure compliance regulation in some countries, which goes beyond the guidelines set out by the International Commission on Non-Ionizing Radiation Protection (ICNIR) [1], [2]. Furthermore, when it comes to 5G BS deployment, operators are facing un-harmonised RF-EMF regulatory challenges in certain countries, regions and even cities. For example, in Switzerland and Italy, a different regulation is put in place where the current RF-EMF exposure limits are 4 V/m and 6 V/m, respectively, which is much stricter than the ICNIRP guidelines at 61 V/m [3].

This restrictive regulation will affect the coverage and quality of the service provided to the consumers and have already had impacts on fourth generation (4G) networks [4] and is envisaged to be problematic for the design and deployment of effective 5G networks [2], [5]–[7] – meaning that 5G communication systems may not work in some geographical areas and that the seamless connectivity promised by 5G may not be possible. Regulators, operators, 5G BS manufacturers, and equipment suppliers all require reliable and agreed assessment of RF-EMF levels exposure to support consistent and effective 5G regulation and network design.

The conventional RF-EMF exposure measurement methods for defining the BS exclusion zone (a compliance boundary around the BS with no access to general public) in both the third and fourth generations, i.e. 3G and 4G, are based on the use of the maximum worst case exposure in every possible direction for a defined time-period. These methods are becoming obsolete to quantify the RF-EMF exposure of 5G BSs employing complex beamforming technology such as massive multiple-input-multiple-output (mMIMO), which allow energy to be focused in sharp high-gain beams in the direction of a specific mobile user.

Different contributions have demonstrated that these methods are not suitable for 5G BSs [8], [9]. Therefore, there is a need to develop new reliable and robust methods for assessing the RF-EMF exposure levels of 5G for supporting consistent and effective 5G regulation and network design; methods based on rigorous scientific evidence to ensure a good balance between user experience and public safety, e.g. ensuring that high power user service beams are only transmitted on a need-to basis [4], [9]. This paper extends our preliminary work in [10] by investigating the stochastic nature of mMIMO 5G beamforming system, based on traceable experimental-based RF-EMF exposure evidence within a real-world indoor environment. The measured data are statistically assessed by evaluating the spatial variations of RF-EMF surrounding the mMIMO testbed with varying beam profiles (i.e. number and position of the different users) and data rates. This paper is organized as follows: Section II describes the experimental setups, Section III presents some measurement results and explains how these results are statistically assessed. Conclusions are drawn in Section IV.

## II. EXPERIMENTAL SETUP

In this paper, a user-programmable mMIMO testbed (which consists of a synchronization unit, a time distribution system, RF transceiver modules). This testbed is capable of generating transmit waveforms for 128 RF channels and independently driving each element in the 128-element transmit antenna array. The system also includes several user equipment (UE) devices each with a 4-element receive antenna array and four independent receivers. A separate RF-EMF acquisition system was also employed which consisted of a triaxial isotropic field probe, Agos ARIA-6000 and a handheld portable spectrum analyser, Keysight FieldFox model N9917B. All the relevant hardware details have been described in [10]. The following provides further details of the measurement setups.

### A. Traceable calibration in laboratory environments

In preparation for the traceable measurement campaigns, the 4-element receive antenna arrays, RF power for the modulated signals of the mMIMO testbed and RF-EMF acquisition system were calibrated within a laboratory environment at the UK National Physical Laboratory (NPL). The antenna calibration measurements were made in the NPL screen fully anechoic chamber (FAC) facility as shown in Fig. 1, which has a dimension of 7 m ×6.2 m ×6.2 m [11]. The four-element receive antenna arrays were calibrated by three-antenna method whereby the relevant cable loss was measured using a Keysight PNA-X model N5242A vector network analyser. The traceable RF power for the modulated signals received by the UE receiver were calibrated with a NI-5681 power sensor. Fig. 2 shows the calibrated E-field results acquired by different receivers for modulated signal centred at 2.63 GHz with 38.88 MHz instantaneous data bandwidth. Note that the theoretical calculation and a calibrated commercial Narda probe were used to validate the obtained calibrated results. These traceable results enable assured confidence over the real-world measurement campaigns. To achieve traceability, this acquisition system has been calibrated at the Power Flux Density Laboratory in UK National Physical Laboratory (NPL) against known E-field [12] (see Fig. 3).

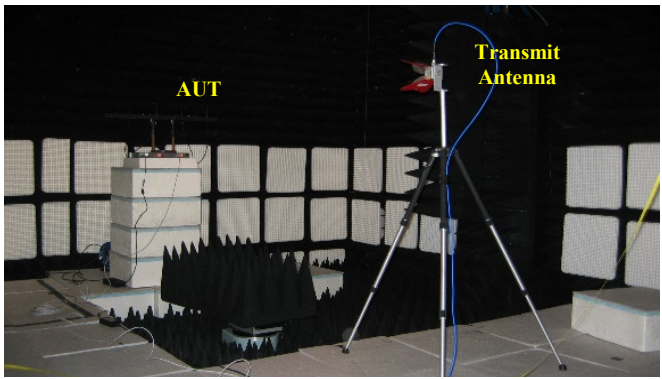

Fig. 1. Measurement Setup for antenna calibration in NPL's fully anechoic chamber.

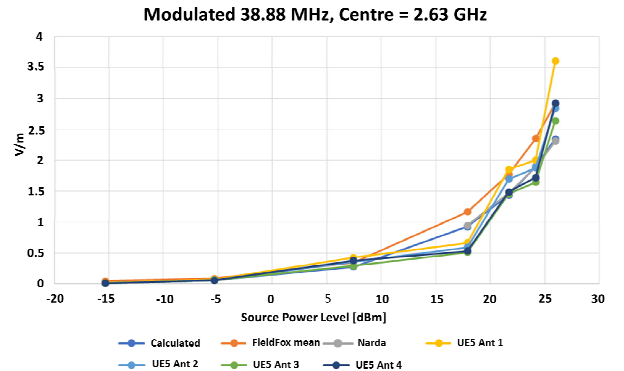

Fig. 2. Calibrated traceable E-field results acquired by different receivers for modulated signal central at 2.63 GHz with 38.88 MHz data bandwidth.

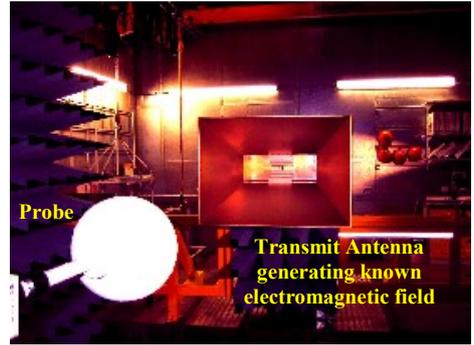

Fig. 3. Measurement setup for field probe calibration in NPL's power flux density laboratory.

### B. Experimental setup in a real-world indoor environment

The real-world measurement campaigns were performed indoors within a large meeting room, located in the basement of the 5G Innovation Centre (5GIC) at the University of Surrey [12]. The room is surrounded by glass, wooden partitions, brick and plasterboard walls and has a dimension of 15 m ×7.5 m ×3 m. The floor is made of concrete and carpeted whereby the concrete ceiling was equipped with some hanging lighting and projector equipment. During the measurements, the typical furniture within the room were placed aside and all the measurement instruments were positioned inside the room (see Fig. 4).

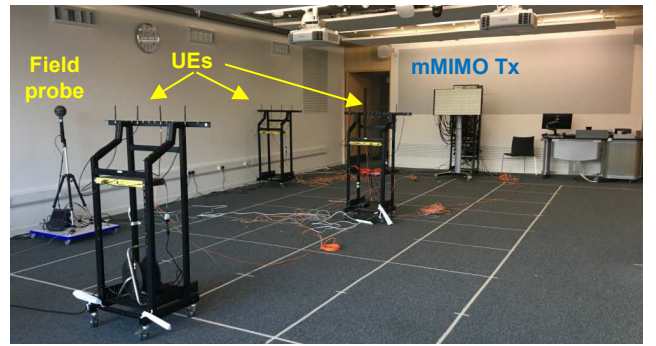

Fig. 4. Measurement Setup for RF-EMF assessment from a mMIMO testbed in an indoor environment.

The mMIMO testbed was programmed to operate with 96 active transmitting antennas and each UE was operated with two vertically polarized dipole receiving antenna. The RF-EMF measurement system was located on a trolley during the measurement (see Fig. 4). During the measurement campaigns, various single-user (SU) and multiple-user (MU) MIMO downlink communication beamforming scenarios were considered, with different combinations of active beams and data traffic loading to mimic the performance of a realistic 5G BS.

The 5G NR beamforming baseband waveform signals were generated by using Keysight PathWave System Design platform (also known as SystemVue). Each generated beam contains a PDSCH (physical downlink shared channel) with random payload data for about 1 millisecond of transmission period. In this paper, up to 4 simultaneous active beams (each transmit a separate IQ waveform signal) were considered, chosen from a total of 19 user-defined potential azimuthal beamforming directions equally spaced every 5° between ± 45° in front of the 128-element mMIMO transmit antenna array. Each element at the mMIMO transmit antenna array is individually driven to form the resultant beam at user-defined targeted azimuthal beamforming directions. Furthermore, the data rate per beam were controlled by changing the modulation coding scheme (MCS, between QPSK (quadrature phase shift keying), 16QAM (16 quadrature amplitude modulation) and 64QAM) and the number of allocated resource block (NRB) used in the transmission.

A total of nine UE and probe setup configurations were considered to represent different acquisition setups and their coordination details (unit in meter) are given in Table 1. Fig. 5 illustrates two configuration setup examples for Configurations 1 and 3. The measurements were conducted at 2.63 GHz with a 38.88 MHz instantaneous data bandwidth per channel and up to 864 total NRB (TNRB) for 4 simultaneous active beams. System control and data processing are performed in Matlab.

TABLE I. COORDINATES OF UE AND PROBE SETUP CONFIGURATIONS

| C[a] | UE1[b] | UE2[b] | UE3[b] | UE4[b] | UE5[b] | Probe[b] |
|---|---|---|---|---|---|---|
| 1 | (-1.5, 3) | (-0.5, 3) | (0.5, 3) | (1.5, 3) | (2.5, 3) | (0, 4) |
| 2 | (-1.5, 3) | (-0.5, 5) | (0.5, 6) | (1.5, 4) | (2.5, 2) | (0, 4) |
| 3 | (-1.5, 7) | (-0.5, 3) | (0.5, 4) | (1.5, 6) | (2.5, 5) | (1, 2) |
| 4 | (-1.5, 5) | (-0.5, 6) | (0.5, 1) | (1.5, 3) | (2.5, 7) | (-1, 3) |
| 5 | (-1.5, 2) | (-0.5, 4) | (0.5, 7) | (1.5, 2) | (2.5, 4) | (1, 5) |
| 6 | (-1.5, 1) | (-0.5, 7) | (0.5, 5) | (1.5, 7) | (2.5, 6) | (-2, 6) |
| 7 | (-1.5, 4) | (-0.5, 8) | (0.5, 2) | (1.5, 5) | (2.5, 8) | (1, 7) |
| 8 | (-1.5, 6) | (-0.5, 1) | (0.5, 8) | (1.5, 1) | (2.5, 1) | (-2, 1) |
| 9 | (-1.5, 8) | (-0.5, 6) | (0.5, 4) | (1.5, 8) | (2.5, 2) | (0, 1) |

a. Configurations
b. Coordinate unit in meter

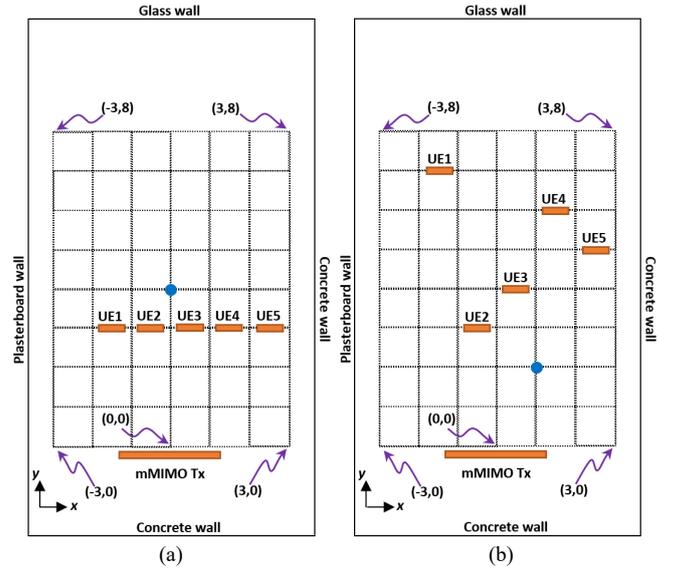

Fig. 5. Measurement setup for 96 active mMIMO Tx at (0 m,0 m): (a) Configuration 1; (b) Configuration 3.

### III. EXPERIMENTAL RESULTS

A total of 228,872 E-field measurements were performed from the 9 configurations in the measurement campaign. The acquired results were organised in form of a large matrix containing parameters used to generate the beam and the measured electric field for each considered transmission scenario with random payload data. By evaluating the measured spatial and temporal variation of RF-EMF surrounding the mMIMO testbed with varying beam profiles and data rates, the following sub-sections provide insights into the stochastic nature of 5G NR mMIMO system in door environment.

#### A. RF-EMF variation at a fixed location

Fig. 6 shows the calibrated RF-EMF for the mMIMO system operating with a single beam steered to the 19 potential azimuthal beamforming directions between ± 45° (with step of 5°) in front of the 128-element mMIMO transmit antenna array, were data are, acquired by using, respectively, the field probe setup under Configuration 3 (Fig. 6(a)), and the 2-element antenna array of UE2 and UE4 setup under Configuration 1 (Figs. 6(b) and 6(c)). Fig. 7 presents the empirical cumulative distribution function (CDF) of the calibrated RF-EMF acquired using the field probe at the same location under Configuration 3 with different number of simultaneous active beams from the mMIMO testbed where each beam steered randomly to one of the 19 user-defined potential azimuthal beamforming directions. From the results shown in Fig. 6, one can observe that the acquired RF-EMF for the mMIMO system operating with a single beam reach its peak at the angle where the receiver is located. The amplitude of RF-EMF is greater in Fig. 6(a) as a result of the receiver being located closer to the mMIMO transmit array. The results in Fig. 7 show that as more beams are simultaneously activated the probability of higher value of RF-EMF is increased.

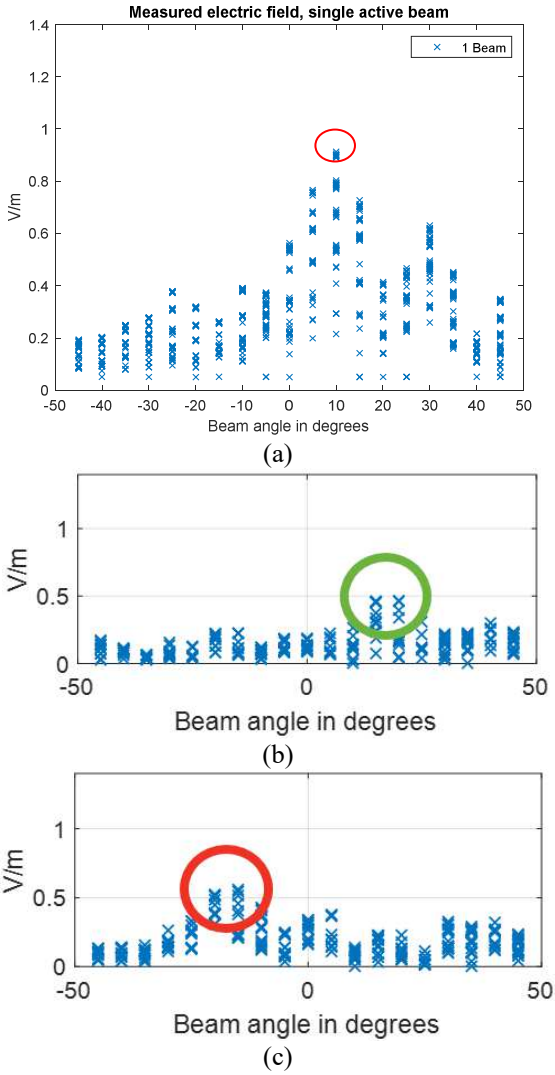

Fig. 6. The calibrated RF-EMF for mMIMO operating with a single beam acquired using: (a) field probe setup under Configuration 3; (b) 2-element antenna array of UE2 setup under Configuration 1; (c) 2-element antenna array of UE4 setup under Configuration 1.

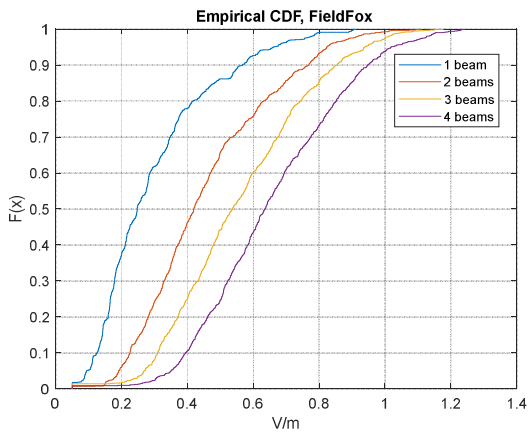

Fig. 7. Empirical CDF of the calibrated RF-EMF acquired using the field probe at the same location under Configuration 3 with different number of simultaneous active beams from the mMIMO testbed.

### B. Effects of Beam Numbers to RF-EMF variation

Fig. 8 shows the calibrated RF-EMF vs distance from the mMIMO testbed for all indoor measurement results with different number of simultaneous active beams when taking into account all different beam directions and data rates. As depicted in Figs. 8(a) and 8(b), the comparison of the acquired results using the Surrey UE receivers (red-cross) and Keysight's RF-EMF probe (blue-cross) shows similar RF-EMF variation trends while they are positioned at similar distance away from the mMIMO. Similar to the observation in Fig. 7, higher values of RF-EMF are obtained when more beams are simultaneously activated.

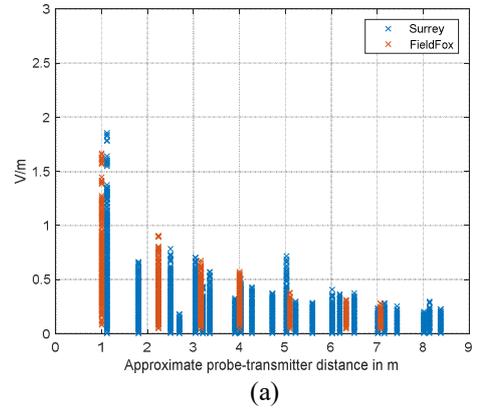

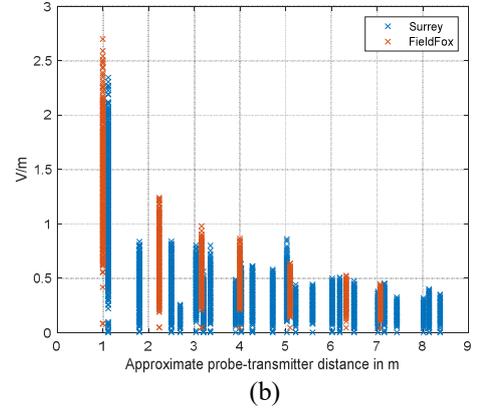

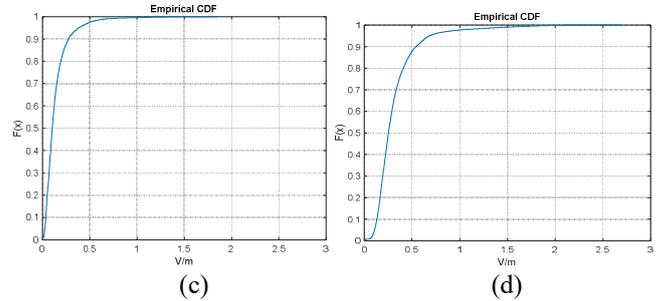

Fig. 8. The calibrated RF-EMF for all indoor measurement results with different number of simultaneous active beams from the mMIMO testbed: (a) 1 active beam; (b) 4 active beam; (c) empirical CDF for 1 active beam; (d) empirical CDF for 4 active beam.

## C. Effects of Data Rate to RF-EMF variation

Fig. 9 shows the calibrated RF-EMF vs distance from the mMIMO testbed for all indoor measurement results with different data rate per beam with different NRB used in the transmission when taking into account all different direction and number of simultaneous active beams. As depicted in Figs. 9(a) and 9(b), the comparison of the acquired results using the Surrey UE receivers and Keysight's probe shows similar RF-EMF variation trends while they are positioned at similar distance away from the mMIMO. The results show that using more NRB implies higher values of RF-EMF.

## IV. CONCLUSIONS

This paper has presented an experimental-based RF-EMF exposure evidence within a real-world indoor environment. The mMIMO tested has up to 128 RF channels with user-programmable SDR capability. The stochastic nature of 5G NR mMIMO system has been statistically assessed by evaluating the measured spatial variation of RF-EMF surrounding the mMIMO testbed with varying beam profiles and data rates. Several other varying factors that influence the RF-EMF of mMIMO system, e.g. user position, have also being considered.

ACKNOWLEDGEMENT

This work was supported by the EU project 5GRFEX entitled – 'Metrology for RF exposure from Massive MIMO 5G base station: Impact on 5G network deployment' (this project has received funding from the support for impact (SIP) programme co-financed by the Participating States and from the European Union's Horizon 2020 research and innovation programme), under European Association of National Metrology Institutes (EURAMET) Reference 18SIP02.

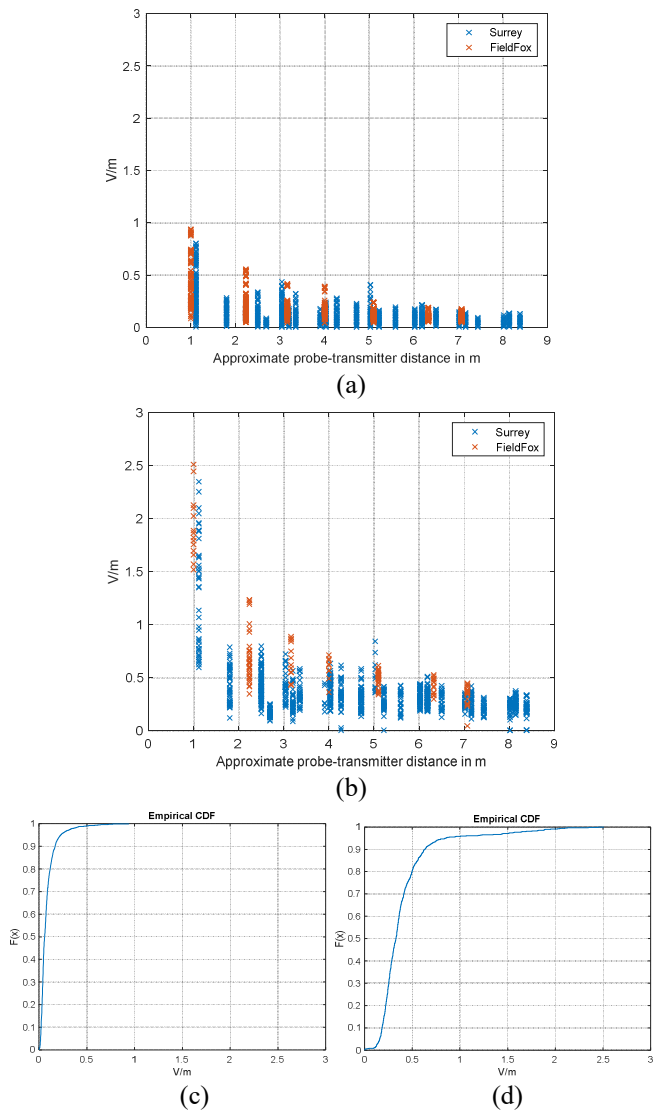

Fig. 9. The calibrated RF-EMF for all indoor measurement results with different data rate per beam with different NRB: (a) 54 resource block (about 6% of TNRB); (b) more than 756 resource block (> 87% of TNRB); (c) empirical CDF for 6% TNRB; (d) empirical CDF for >87% TNRB.